\documentclass[useAMS,usenatbib]{mn2e}

\usepackage{graphicx, gensymb,epsfig}
\usepackage{amsmath}

\pdfoutput=1

\title{Directional Time-Distance Probing of Model Sunspot Atmospheres}

\author[H. Moradi,  P. S.~Cally, D. Przybylski, and S. Shelyag ]{H. Moradi, \thanks{E-mail: hamed.moradi@monash.edu}
P. S.~Cally, D. Przybylski and S. Shelyag\\
Monash Centre for Astrophysics and
School of Mathematical Sciences,
Monash University, Victoria, Australia 3800}

\newcommand{\Science}{Science}

\usepackage{aas_macros}

\begin{document}

%\label{firstpage}

\maketitle

\begin{abstract}

%Local helioseismology is typically used to study surface features on the Sun, such as sunspots. 
A crucial feature not widely accounted for in local helioseismology is that surface magnetic regions actually open a window from the interior into the solar atmosphere, and that the seismic waves leak through this window, reflect high in the atmosphere, and then re-enter the interior to rejoin the seismic wave field normally confined there. In a series of recent numerical studies using translation invariant atmospheres, we utilised a ``directional time-distance helioseismology'' measurement scheme to study the implications of the returning fast and Alfv\'en waves higher up in the solar atmosphere on the seismology at the photosphere \citep{cm2013,mc2014}. In this study, we extend our directional time-distance analysis to more realistic sunspot-like atmospheres to better understand the direct effects of the magnetic field on helioseismic travel-time measurements in sunspots. In line with our previous findings, we uncover a distinct frequency-dependant directional behaviour in the travel-time measurements, consistent with the signatures of MHD mode conversion. We found this to be the case regardless of the sunspot field strength or depth of its Wilson depression. We also isolated and analysed the direct contribution from purely thermal perturbations to the measured travel times, finding that waves propagating in the umbra are much more sensitive to the underlying thermal effects of the sunspot.
\end{abstract}


%==================================================================
\section{Introduction}

Sunspots and active regions (magnetic flux concentrations tens of thousands of kilometres across containing sunspots) are the most visible manifestation of solar magnetic activity on the solar surface. A detailed understanding of sunspots and magnetically active regions is therefore essential in order to establish accurate physical relationships between internal solar properties and magnetic activity in the photosphere. 

Using observations of surface oscillations, helioseismology provides the most effective way to observationally probe structure inside the Sun. The combination of high spatial resolution, continuous observing, and simultaneous vector magnetograms provided by the Helioseismic and Magnetic Imager (HMI) instrument on board the Solar Dynamics Observatory (SDO) delivers unprecedented probing of magnetic active regions and sunspots. 

However, important developments in the techniques of local helioseismology (i.e., both theoretical and observational) are required to realize the full potential that these observations offer. As highlighted by a number of detailed comparative studies and reviews on the matter \citep[e.g.,][]{gizonetal2009,moradietal2010,moradi2012}, major challenges exist in the development of new helioseismic procedures that are robust in the presence of magnetism and capable of probing both subsurface magnetic structures and associated flows.  

Over the years various local helioseismic techniques have substantially contributed to our understanding of the solar interior \citep[see][for a comprehensive review]{gbs2010}. The most widely used measurement method in local helioseismology is time-distance helioseismology \citep{duvalletal1993}. By correlating observations of Doppler velocity at different times and positions on the solar surface, a causative link is inferred and a travel time between pairs of points putatively determined. Comparing these travel times with those calculated for the quiet-Sun, one infers the presence of wave-speed anomalies beneath the surface that may be due to such features as magnetic fields, temperature variations or plasma flows. 

With the noise level being substantially larger for group-time measurements  \citep{kds2000}, time-distance travel times are typically derived from phase travel-times. While this is adequate for the quiet Sun, interpretation becomes more complicated when considering active regions, where phase shifts can naturally arise from changes in wave propagation speed  (e.g., due to subsurface flows and sound speed perturbations induced by the presence of the magnetic field), but they can also result from other sources as well. For example, mode damping \citep{woodard1997}, acoustic source suppression \citep{gb2002}, and the Wilson depression \citep{bs2000,lb2000} have all been identified as possible sources of phase shifts in active regions. 

Another important source of phase shifts is via MHD mode conversion in the atmosphere. Mode conversion takes place in regions where the sound ($c_s$) and Alfv\'en speed ($c_a$) are comparable. It is expected to be significant for sunspot seismology because in the umbrae of sunspots, the layer where $c_a = c_s$ lies is just a few hundred kilometres below the formation height of the Fe I spectral line of the Helioseismic and Magnetic Imager (HMI) instrument on board the Solar Dynamics Observatory (SDO) . 

``Fast- to-slow'' mode conversion has been explored as a primary cause of acoustic wave ($p$-mode) absorption in sunspots for decades \citep{sb1992,cbz1994,
cally1995,cc2003,shelyagetal2009}. In this scenario, the $p-$modes emerging in sunspots below the $c_a = c_s$ level are effectively (acoustic) fast waves. On passing through the $c_a = c_s$ layer the $p$-modes are partially transmitted into the solar atmosphere as (primarily acoustic) slow waves, most efficiently at small ``attack angle'' (the angle between the wavevector and the magnetic field). The transmitted sound waves propagate longitudinally along field lines at frequencies above the field-adjusted acoustic cutoff frequency and are reflected otherwise \citep{bl1977}.  If the attack angle is not small however, significant amounts of energy will be converted to (magnetic) fast waves \citep{sc2006}. These fast waves are then reflected off the Alfv\'en wave speed gradient, at the height where their horizontal phase speed ($v_{ph} = \omega/k_h$; where $\omega$ is the angular frequency and $k_h$ the horizontal wavenumber) is approximately equal to $c_a$, back down to the surface (having assumed $c_a\gg c_s$ at this level). 

More recently it has been realised that fast waves created in this way are further subject to partial conversion to Alfv\'en waves higher in the atmosphere. Depending on the local relative inclinations and orientations of the background magnetic field and the wavevector, the fast wave may undergo partial mode conversion to either an upward or downward propagating Alfv\'en wave around the reflection height, where they are near-resonant \citep{cg2008,ch2011,kc2011,kc2012,felipe2012}. After they reflect off the Alfv\'en wave speed gradient, the fast waves may re-enter the solar interior wave field. This could be problematic for helioseismology, since any phase changes produced by the ``fast-to-Alfv\'en'' mode conversion process would seriously compromise any inferences derived from helioseismic inversions of phase travel times \citep[e.g.,][]{duvalletal1996,kds2000,couvidatetal2005}, which would normally, but inaccurately, interpret such phase changes as ``travel-time shifts'' due to subsurface inhomogeneities alone.

In a series of recent numerical studies \citep{cm2013,mc2013,mc2014}, we quantified the implications of the returning fast and Alfv\'en waves for the seismology of the photosphere by comparing Alfv\'enic losses higher up in the solar atmosphere with helioseismic travel-time shifts at the surface. Using 3-D numerical simulations of helioseismic wave propagation in simple translationally invariant atmospheres, we applied a ``directional time-distance helioseismology" approach sensitive to magnetic field orientation, finding substantial wave ``travel time'' discrepancies of several tens of seconds (depending on field strength, frequency, and wavenumber) related to phase changes resulting from mode conversion, and not ``actual'' travel time changes. These results, which were also verified using the Boundary Value Problem (BVP) method of \cite{cg2008} and \cite{cally2009b}, indicated that processes occurring higher up in the atmosphere are strongly influencing the core data products of helioseismology. 

In these studies only translation invariant setups were used, which are most useful to study the effect in fundamental terms, but are not typically found on the Sun. Our best chance at constraining the interior structure of sunspots comes with constructing accurate forward models. Hence, in order to be able to make meaningful estimates of the direct role played by MHD mode conversion and wave reflection in helioseismic measurements, we extend our study to more realistic sunspot model atmospheres spanning the subphotosphere ($z=-10$ Mm, with $z = 0$ being the photosphere) to the chromosphere ($z= 1.9$ Mm) and study the sensitivity of directional helioseismology measurements to changes in the photospheric and subsurface structure of sunspot models.  For practical computational reasons, we were unable to model the seismic effects of the Transition Region, though \cite{hc2014} find it too has significant signatures. 

\section{The Background Model}

The background models we employ consist of a number of azimuthally symmetric magneto-hydrostatic (MHS) sunspot atmospheres adopted from \cite{kc2008}. To summarise, these models consist of a concatenation of a self-similar model in the deep photospheric layers, calculated following the method of \cite{low1980}, with a potential solution above some arbitrary height using the method of \cite{pizzo1986}. The thermodynamic variables for the sunspot axis are taken from the semi-empirical \cite{avrett1981} umbral-core model, while the ``quiet-Sun'' atmosphere variables are taken from Model S \citep{cdetal1996} in the deep sub-photospheric layers, smoothly joined to the VAL-C \citep{val1981} model in the photospheric and chromospheric layers, and stabilised using the method outlined by \cite{pk2007}. 

%This stabilisation was performed in a way that minimised the changes to the photospheric pressure and density profiles to allow for realistic radiative diagnostics. 

The sunspot models possess a high degree of flexibility for conducting a detailed directional helioseismology study, with a number of variable parameters such as field strength on the axis, field inclination at the photosphere, and spot radius. Another important variable parameter in the models, is the Wilson depression -- the height difference between the umbra and photosphere. This can be easily changed in the model by choosing the desired location of the constant optical depth $\log \tau_{5000} = 0$ (the formation height of the $5000$ \AA~continuum radiation) of the semi-empirical umbral model. Studies have shown that the Wilson depression may be a signifiant source of travel time reductions in sunspots \citep{bl2000,lcr2010}, so it is important to observe what effects it may have on directional travel-time measurements. 

We conduct a number of experiments with the sunspot models. In the first set of experiments, we study the sensitivity of directional travel times to sunspot field strength and inclination using two sunspots models with differing peak photospheric field strengths (1.5 and 2.5 kG) but with all other parameters fixed (i.e., spot size, field inclination at the photosphere and Wilson depression, which is fixed at 400 km). In the second set of experiments, we investigate the sensitivity of the directional travel times to the depth of the Wilson depression. Estimates from observations put the depth of Wilson depression in the range of $300 - 1500$ km  \citep{bl1964,mpv1993,mathewetal2004,watsonetal2009}. For our study we employ three identical sunspot models with a relatively moderate surface field strength ($1.5$ kG) with varying Wilson depressions depths of $300$, $400$ and $500$ km respectively. In our final set of experiments, we ascertain the contribution of the underlying thermal perturbations to the travel-time shifts in contrast to the direct magnetic effects in a sunspot with surface field strength of $1.5 $ kG and a Wilson depression of 400 km. As shown in \cite{mc2008} and \cite{mhc2009} this can easily be achieved in linear numerical simulations by suppressing the direct magnetic effect on the waves. The combined outcomes from these experiments provide us with valuable diagnostics of both the thermal and magnetic structures of sunspots.
\begin{figure}
\begin{center}
\includegraphics[width=\hsize, trim = 1cm 4cm 1cm 4cm] {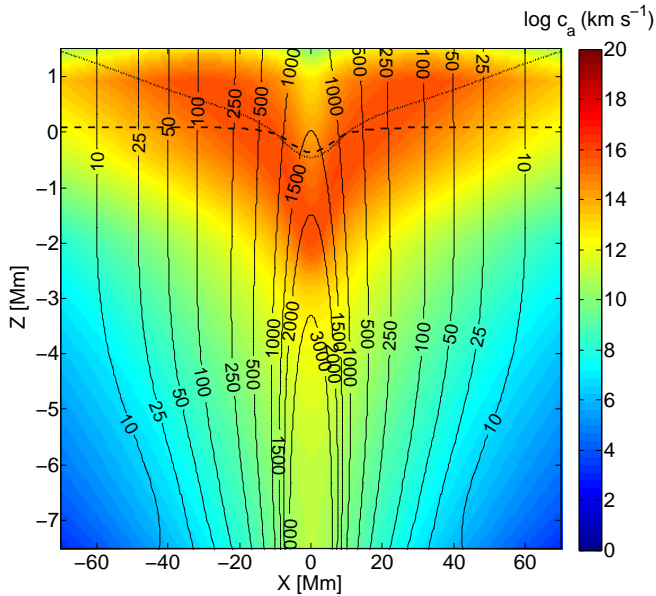}
\caption{An example of the MHS sunspot atmosphere used in the forward modelling. A cut through $y=0$ for a $1.5$ kG sunspot with a Wilson depression of 400 km is shown here. The background colour-scale corresponds to $\log c_a$ (km s$^{-1}$). The contour lines represents the field strength (in G). The dotted line indicates the location of the plasma $\beta \approx 1$ layer, while the dashed line represents the reference observation height corresponding to $\log\tau_{5000} =  - 1.6$. } 
\label{fig:spot_plot}
\end{center}
\end{figure}

As all sunspot models encompass a Wilson depression, the simulated data are analysed at a constant optical depth $\log\tau_{5000} =  - 1.6$, which roughly represents the layer where the contribution function for FeI $6173$ \AA~photospheric spectral line has its maximum \citep{khomenkoetal2009}. The optical depths are calculated using the routine described in \cite{jessetal2012} by integrating the continuum and line opacities along the lines-of-sight for each column in the sunspot models. The ATLAS9 package \citep{kurucz1993} opacities are used in the computation. We also have the ability to choose the line-of-sight viewing angle (from the vertical), but for simplicity we calculate and compare directional travel times using the photospheric velocity at disk centre. Some properties of these sunspot model atmospheres are shown in Figure $\ref{fig:spot_plot}$. 

\section{The Forward Model}

As in our previous studies, we numerically solve the linearised equations of ideal MHD using the Seismic Propagation through Active Regions and Convection (SPARC) code \citep{hanasoge2007} which has been successfully utilised in the past to study wave propagation through model sunspots \citep{mhc2009}. The dimensions of the 3-D computational box employed for the numerical simulations are 140 Mm in the horizontal ($x,y$) directions and 11.9 Mm in the vertical ($z$) direction. The bottom boundary of the domain is located at 10 Mm below the photospheric level $z=0$. The horizontal grid spacing consists of 256 equidistant points in $x$ and $y$, with a resulting resolution of $\Delta x = \Delta y \approx 0.55$ km/pixel, while the vertical grid spacing $\Delta z$ is nonuniform, ranging from tens of kilometres near and above the surface to just over one hundred kilometres near the bottom of the computational domain. The top $\sim500$ and bottom $\sim800$~km of the box are occupied by the vertical absorbing (PML) boundary layers, while absorbing sponges line the sides of the box.      

\begin{figure}
\begin{center}
\includegraphics[width=\hsize,trim = 4cm 2cm 4cm 2cm]{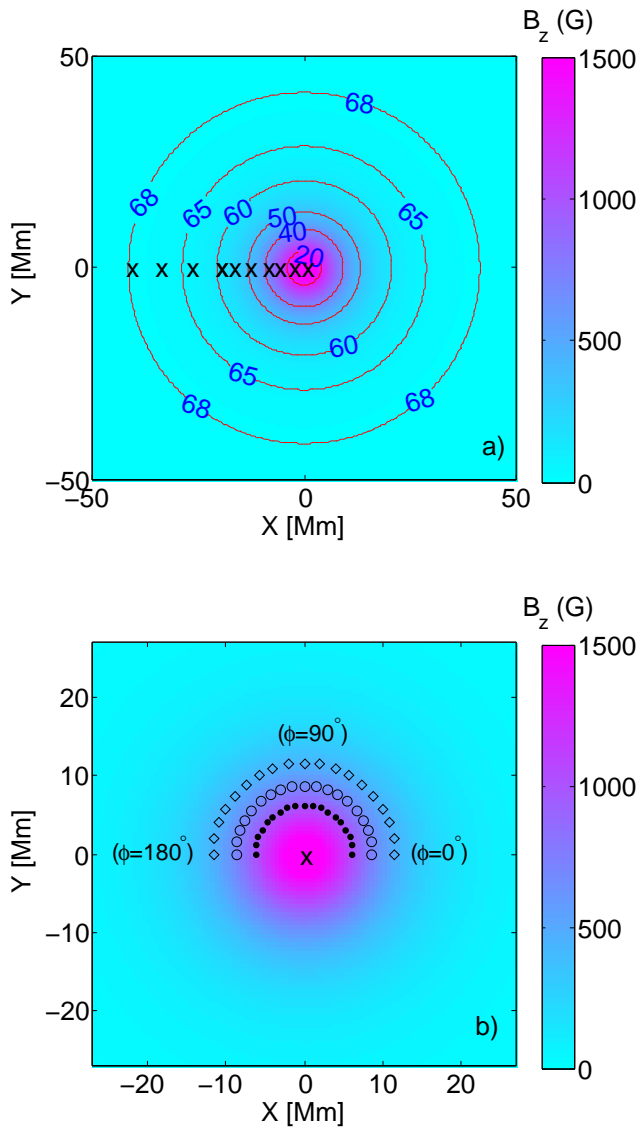}
\caption{The one-way travel time measurement geometry. In both panels, the background represents the vertical component of magnetic field strength at the observation height $\log\tau_{5000}=-1.6$. In panel a) the contours are indicative of the magnetic field inclination from the vertical ($\theta$ in degrees) at the same height in the atmosphere, while the crosses represent the locations of the individual acoustic sources utilised in the forward modelling calculations. Panel b) shows an example of the receiver locations (which span from $0 \le \phi \le 180\degree$, from the right- to the left-hand side of the sunspot, spaced $\phi = 10 \degree$ apart) for a source initiated at $\theta = 0\degree$, denoted by the cross on the axis $(x = 0$, $y = 0$). The dots indicate the receiver locations at $\Delta=6.2$ Mm, the circles $\Delta=8.7$ Mm, and the diamonds represent $\Delta=11.6$ Mm.  } 
\label{fig:sources}
\end{center}
\end{figure}

The axis of the sunspot is placed at the centre of the computational domain. For each sunspot case studied, we conduct ten unique simulations using a Gaussian perturbation source positioned along the left hand side of the sunspot along $y=0$, starting from the axis ($x = 0$, $y = 0$) and then at nine other locations along the negative $x$ axis, as depicted in Figure \ref{fig:sources}. As the sunspot model is axisymmetric, this allows us to study each corresponding $\theta$ associated with the source location separately. 

The acoustic source employed for our calculations is similar to that employed by \cite{shelyagetal2009} and \cite{mc2014}, where a source term of the form: 
\begin{equation}
v_z = \sin \frac{2\pi t}{t_1} \exp \left( -\frac{(r-r_0)^2}{\sigma_r^2} \right) \exp \left( - \frac{(t-t_0)^2} {\sigma_t^2} \right), 
\end{equation}  
is added to the right hand side of the vertical momentum equation. In the equation above $v_z$ is the perturbation to the vertical component of the velocity, $t_0 = 300 $s, $t_1= 300 $s, $\sigma_t = 100 $s, $\sigma_r = 4\Delta x$, and $r_0 (x,y)$ is the source position. The source, which is always initiated below the surface at $z = - 0.65$ Mm, generates a broad spectrum of acoustic waves in the $3.33$ mHz range, mimicking wave excitation in the Sun.  For each magnetic case we conduct a separate quiet-Sun run to act as the reference (unperturbed) model. 

The relatively large field strengths associated with the sunspot models being considered, coupled with the exponential drop in density with height in the atmosphere, naturally results in a substantial $c_a$ above the surface. For explicit numerical solvers, such as SPARC, this results in severe CFL ($\Delta t \approx \Delta z /c_a$) constraints, significantly compounding the computational expense of conducting a detailed parametric study. To alleviate the problem, we employ a Lorentz Force ``limiter'' to limit/cap the Alfv\'en wave speed at a particular value above the surface. This approach is commonly adopted by explicit numerical solvers in computational MHD studies of sunspot structure \citep{rsk2009,cameronetal2011,braunetal2012}, allowing one to increase the simulation $\Delta t$ to any desired or practical value. 

However in \cite{mc2014}, we studied the physical implications of imposing an artificial limit on $c_a$ and found that it can severely impact the fast-wave reflection height ($c_a \approx \omega/k_h$) in the sunspot atmosphere, which can be problematic for fast-to-Alfv\'en mode conversion and any subsequent helioseismic analyses. In fact, we found that unless the $c_a$ cap is placed well above the horizontal phase speed associated with the wave travel distance being studied (thus ensuring minimal damage to the fast-wave reflection height), helioseismic travel time measurements could be severely affected. On the back of these findings, we decided to employ a limiter with a $c_a$ cap at $80$ km s$^{-1}$, but restrict our helioseismic analyses to waves with horizontal phase speeds well below this (see Table \ref{Tab:Delta}), so as to ensure our travel time measurements would not be compromised. 

\section{Directional Time-Distance Helioseismology}

With single source wave excitation, time-distance diagrams can easily be constructed by plotting the resulting velocity signal as functions of time for all horizontal locations. Moreover, each source location along the negative $x$-axis corresponds to a specific field inclination $\theta$ (from the vertical). As seen in Figure $\ref{fig:sources}$, with the source locations we have chosen we can sample $\theta$ in the range $0\degree-70\degree$. By selecting a receiver location at a horizontal distance ($\Delta$) away from the source location around the $xy$-plane, we isolate the magnetic field orientation with respect to the vertical plane of wave propagation, which we refer to as the ``azimuthal'' field angle ($\phi$, where $0 \le \phi \le 180\degree$, from the right- to the left-hand side of the sunspot) which we sample in $10\degree$ bins. 

Prior to calculating the travel times, we first filter the data cubes in two frequency ranges: $3$ and $5$ mHz by employing a Gaussian frequency filter with a dispersion of $0.5$ mHz. We also apply an $f$-mode filter to remove the contribution from surface gravity waves. We then measure the phase travel time perturbations $\delta\tau$ (i.e., the differences in the phase travel times between the magnetic and nonmagnetic simulations) using Gabor wavelet fits \citep{kd1997} to the time-distance diagram at various $\Delta$ away from the source, for each source ($\theta$) receiver ($\phi$) pair of points. A rectangular window of width $14$ minutes centred on the first-bounce ridge selects the fitting interval in time lag. The fits are done by minimising the misfit between the Gabor wavelet and the wave form. An initial guess of the Gabor wavelet parameter values is obtained by fitting the reference (quiet-Sun) wave form first. We use MATLAB's multidimensional unconstrained nonlinear minimisation routine \emph{fminsearch} for the fitting, which employs the Nelder-Mead simplex algorithm \citep{lagariasetal1998}. This is a direct search method that does not use numerical or analytic gradients. We measured $\delta\tau$ for three typical skip distances $\Delta$ \citep{couvidatetal2005}. The horizontal phase speeds associated with these distances are presented in Table \ref{Tab:Delta}.

\begin{table}
 \caption{Wave travel distances analysed and their associated horizontal phase speeds in the quiet Sun model.}
 \label{symbols}
 \begin{tabular}{@{}lcc}
  \hline
  $\Delta$ (Mm) & $v_{ph}$ (km s$^{-1}$) \\
  \hline
  6.2 & 12.6   \\
  8.7 & 14.1  \\
  11.6 & 16.4 \\
  \hline
 \end{tabular}\label{Tab:Delta}
\end{table}
 
\section{Results \& Analysis}

\subsection{Sensitivity of Directional Travel Times to Frequency, Field Strength and Inclination}\label{sec1}

\begin{figure}
\begin{center}
\includegraphics[width=\hsize, trim = 2cm 1cm 2cm 0cm]{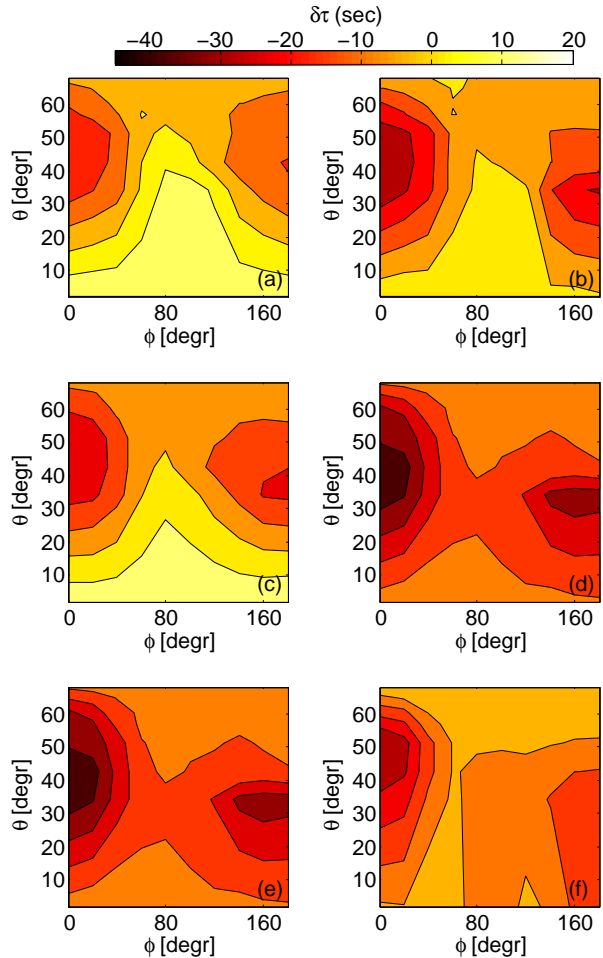}
\caption{One-way phase travel-time perturbations ($\delta\tau$) derived from the $1.5$ kG sunspot model calculations as a function of field inclination ($\theta$) from the vertical, and azimuthal angle ($\phi$) for wave travel distances of $\Delta = 6.2$ (a-b), $8.7$ (c-d) and $11.6$ (e-f) Mm. Left column represents 3 mHz and right column 5 mHz. } 
\label{fig:1p5kG_times}
\end{center}
\end{figure}

\begin{figure}
\begin{center}
\includegraphics[width=\hsize, trim = 2cm 1cm 2cm 0cm]{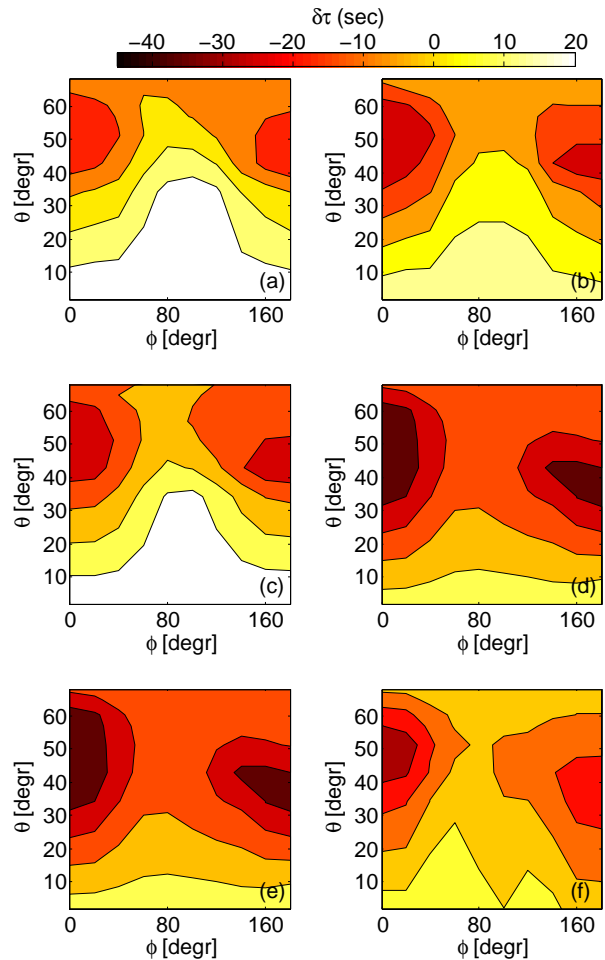}
\caption{Same as Figure $\ref{fig:1p5kG_times}$ but for the $2.5 $ kG sunspot model.} 
\label{fig:2p5kG_times}
\end{center}
\end{figure}

The contour plots in Figures $\ref{fig:1p5kG_times}$ and $\ref{fig:2p5kG_times}$ depict the time-distance phase travel-time perturbations (with respect to the quiet solar model) as functions of wave source position/field inclination from vertical $\theta$ and receiver location/azimuthal direction $\phi$, derived from sunspot models with surface field strengths of $1.5$ and $2.5$ kG respectively. The results shown are for the two frequency bands analysed, 3 (left column) and 5 (right column) mHz, and for waves which travel a horizontal distance of $\Delta = 6.2$ (panels a-b), $8.7$ (panels c-d) and $11.6$ Mm (panels e-f) from the source. 

Pleasingly, the features we observe in Figures $\ref{fig:1p5kG_times} - \ref{fig:2p5kG_times}$ are very much in accord with the directional travel times derived from previous studies using simple translationally invariant background atmospheres \citep{cm2013,mc2014}. Specifically, they show a clear manifestation of the acoustic cutoff at $\theta = 30\degree-40\degree$ for 5 mHz and $\theta = 50\degree-60\degree$ at 3 mHz. For $\theta$ below the acoustic cutoff, small positive $\delta\tau$ values of a few seconds are apparent.  For larger $\theta$ (i.e, sufficient for the ramp effect to take hold $\omega > \omega_c \cos \theta$), the atmosphere is open to wave penetration and mode conversion. This results in significant negative $\delta\tau$ for these $\theta$, particularly at small sin $\phi$ (around $0\degree$ and $180 \degree$), which is due to the fast magnetically-dominated waves undergoing significant phase enhancement on returning to the surface after passing upward through the $c_a = c_s$ layer, reflecting near $\omega/k_h = c_a$, and finally re-entering the interior via $c_a = c_s$ again. However, away from $\phi=0$ (and $180\degree$), the fast waves lose energy as they are partially converted to the Alfv\'en wave, which results in a phase retardation that partially cancels the underlying negative travel time perturbation at small $\sin\phi$. In line with previous studies, the energy loss is at its maximum around $\phi=80\degree-100\degree$, orientations typically associated with peak fast-to-Alfv\'en conversion \citep{kc2011,kc2012,cm2013}.  

However unlike our previous studies where the atmosphere and magnetic field were horizontally invariant, the presence of the sunspot, coupled with the distribution of the individual wave sources (on the left-hand side of the sunspot), results in a distinct asymmetry in $\delta\tau$ about $\phi$. This is essentially due to one end of the wave path being in a stronger region of perturbation (i.e., inside the sunspot), and the other end being near or inside the ``quiet Sun'' region, which will naturally result in a directional bias in $\delta\tau$, with larger (negative) $\delta\tau$ expected for $\phi < 50 \degree$ (i.e., waves travelling primarily to the right/inside the ``umbra'' of our sunspot model). This effect is exacerbated as the wave travel distance is increased, as is evident in Figures $\ref{fig:1p5kG_times}- \ref{fig:2p5kG_times}$ e) and f), for $\Delta = 11.6$ Mm. 

As expected, the magnitude of the $\delta\tau$ perturbations is also strongly dependant on frequency and magnetic field strength. Larger negative and smaller positive $\delta\tau$ are observed for all sunspot models and $\Delta$ as the frequency is increased from 3 to 5 mHz. This frequency dependance of helioseismic travel times has been well documented in the past \citep{bl2000,chou2000,bb2006,cr2007,mhc2009}. Increasing the field strength of the sunspot naturally shifts the location of the $c_a = c_s$ layer deeper below the surface, but the only direct effect on the directional $\delta\tau$ we observe at the photosphere is an increase in their magnitude at both 3 and 5 mHz for all $\Delta$.      

\subsection{The Effect of the Wilson Depression}

\begin{figure}
\centering
\includegraphics[width=\hsize, trim = 2cm 1cm 2cm 0cm] {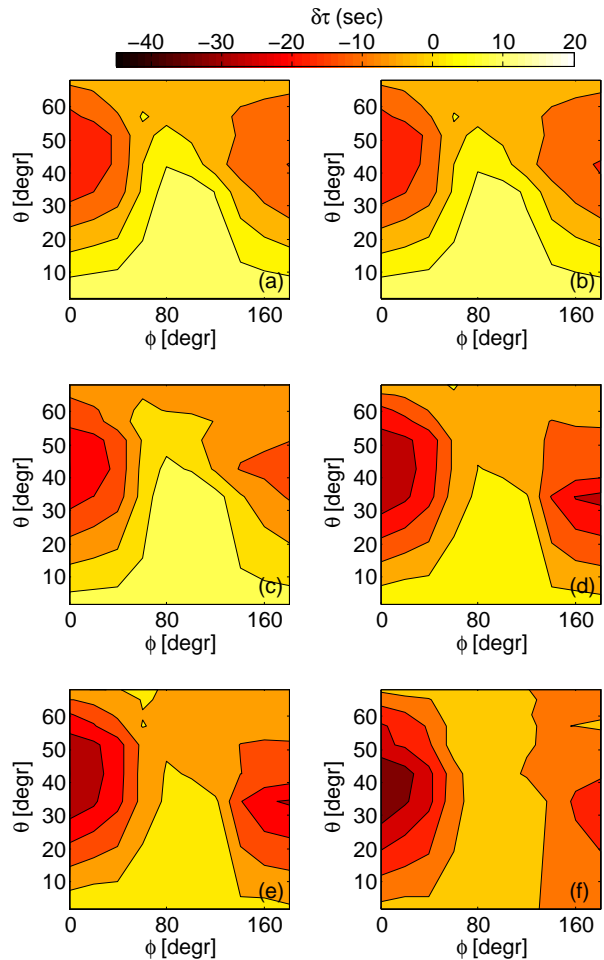}
\caption{One-way phase travel time perturbations $\delta\tau$  as a function of field inclination from the vertical $\theta$ azimuthal angle $\phi$ for $\Delta=6.2$ Mm, derived from three $1 .5$ kG sunspot models with Wilson depression of $300$ (a-b), $400$ (c-d) and $500$ (e-f) km. Left column represents $3$ mHz and right column $5$ mHz. } 
\label{fig:wd_times_d1}
\end{figure}

Figure $\ref{fig:wd_times_d1}$ shows the directional $\delta\tau$ derived for $\Delta=6.2$ Mm for three $1.5$ kG sunspot models with varying Wilson depression depths ($300$, $400$ and $500$ km), calculated at $3$ and $5$ mHz. While the general behaviour of $\delta\tau$ across $\theta$ and $\phi$ for all three models is consistent with those derived in section \ref{sec1}, it is also apparent that modifying the depth of the Wilson depression can have a direct and measurable impact on the directional travel times. 

This is not entirely unexpected of course, as the Wilson depression is a physical displacement in the photosphere which will naturally give rise to a change in the path length of the waves. Modifying the depth of the Wilson depression also implies a change in the near surface density and temperature stratification of the sunspot, which in turn will also modify the actual wave speed (both $c_s$ and $c_a$), the result of which should manifest itself in the travel time calculations. 

\begin{figure}
\centering
\includegraphics[width=1.0\hsize, trim = 1cm 9cm 1cm 9cm]{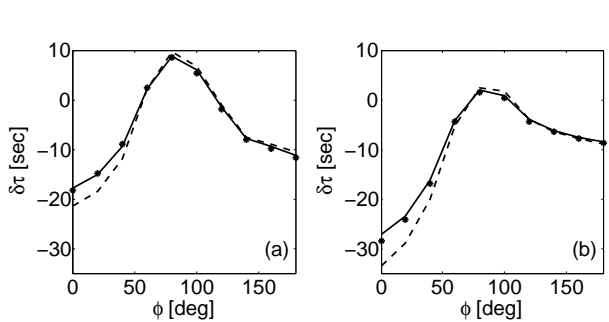}
\caption{One-way phase travel time perturbations $\delta\tau$ for waves initiated at $\theta \approx 32\degree$ as a function of azimuthal angle for $\Delta=6.2$ Mm, derived from three $1 .5$ kG sunspot models with varying Wilson depression depths. The solid lines are the results from $300$ km model, the dotted lines represents the $400$ km model, and the dashed lines represent the $500$ km model. Panel a) represents $3$ mHz travel times and panel b) $5$ mHz travel times. } 
\label{fig:wd_plot}
\end{figure}

These effects are evident in Figure \ref{fig:wd_plot}, which shows a cut at $\theta \approx 32\degree$ through Figure $\ref{fig:wd_times_d1}$. Here we can clearly observe faster travel times associated with waves travelling towards the sunspot axis as the Wilson depression is shifted deeper below the surface. While the $\delta\tau$ differences between $300$ and $400$ km are very subtle (under $\sim 1$ second), more significant differences in $\delta\tau$ are observed with the Wilson depression at $500$ km. At $\phi=0$ we see a $\sim 3$ second difference to the $300-400$ km cases, and at $5$ mHz it's $\sim 7$ seconds. Waves travelling away from spot centre (large $\phi$) do not appear to be affected by the change in Wilson depression depth. We observed a similar behaviour for $\Delta = 8.7$ and $11.7$ Mm (but have not shown them here for the sake of brevity). 

These results are generally consistent with the recent findings of \cite{schunkeretal2013}, who studied the sensitivity of helioseismic travel times to the depth of the Wilson depression using numerical forward modelling of plane wave packets through non-MHS sunspot model atmospheres. They found that a $\sim50$ km change in the Wilson depression can be detected above the observational noise level. 

\subsection{The Effect of Thermal Perturbations}

\begin{figure}
\centering
\includegraphics[width=\hsize,  trim = 2cm 1cm 2cm 0cm]{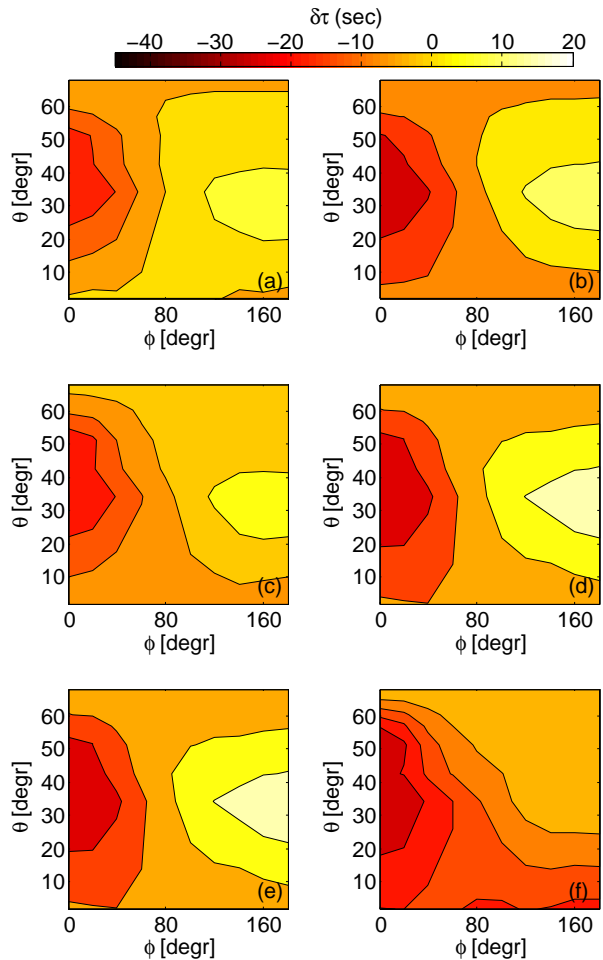}
\caption{One-way phase travel time perturbations ($\delta\tau$) derived from a model where only the thermal perturbations corresponding to the $1.5$ kG sunspot with a Wilson depression of $400$ km are present. Panels (a-b) represent a wave travel distance of $\Delta = 6.2$ Mm, (c-d) represent $\Delta = 8.7$ Mm, and (e-f) represent $\Delta = 11.6$ Mm. Left column represents $3$ mHz travel times and right column $5$ mHz. } 
\label{fig:1p5kG_thermal_times}
\end{figure}

\begin{figure}
\centering
\includegraphics[width=\hsize,  trim = 1cm 9cm 1cm 9cm]{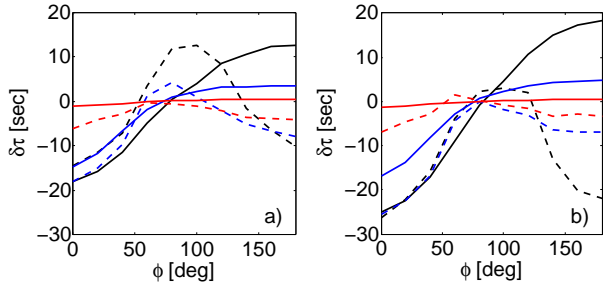}
\caption{Line-plots of one-way phase travel time perturbations $\delta\tau$ along $\phi$ for waves initiated at $\theta \approx 34\degree$ (black), $\theta \approx 51\degree$ (blue) and $\theta \approx 65\degree$ (red) for $\Delta=6.2$ Mm. The solid lines denote travel times derived from the $1 .5$ kG thermal sunspot model, while the dashed lines denote the magnetic sunspot model. Panel a) represents $3$ mHz travel times and panel b) the $5$ mHz travel times. } 
\label{fig:1p5kG_thermal_times_plot}
\end{figure}

One of the advantages of forward modelling of waves in a model sunspot atmosphere is that it provides us with the opportunity to isolate the individual effects of the magnetic field and thermal perturbations on travel-time measurements \citep{mc2008,mhc2009}. In order to isolate the thermal contributions to the measured directional $\delta\tau$, we repeat our single source calculations, this time using a ``thermal" sunspot model, where only the thermal perturbations corresponding to the $1.5$ kG  sunspot model with a Wilson depression of $400$ km are present, but with the direct magnetic effects on the waves suppressed. The directional travel times are then measured in an identical manner as before, with the results shown in Figure $\ref{fig:1p5kG_thermal_times}$. These travel times can be directly compared with those derived from the $1.5$ kG magnetic sunspot in Figure $\ref{fig:1p5kG_times}$, which we do so in Figure $\ref{fig:1p5kG_thermal_times_plot}$, where we show some line plots with the thermal and magnetic travel times plotted on the same scale for a selection of $\theta$. 

It is important to note that the resulting ``thermal travel-time perturbations''  produced from these calculations result from a combination of thermal perturbations and geometrical effects due to the presence of a Wilson depression. When considering purely thermal effects on their own, i.e, a cooler plasma with a reduced sound speed and no Wilson depression, one would expect to see positive travel-time shifts with respect to the quiet Sun, as waves travel slower in the cooler medium. On the other hand, the presence of a Wilson depression can change the wave-path length, depending on the wave propagation direction (towards or way from the umbra for example) and frequency. 

In Figure $\ref{fig:1p5kG_thermal_times}-\ref{fig:1p5kG_thermal_times_plot}$ it is clearly evident that the geometrical effects introduced by the Wilson depression are indeed significant for waves travelling towards the ``umbra'' ($\phi < 50 \degree$), with the reduction in path length seemingly overriding the effects of the cooler plasma, resulting in similar (negative) travel-time shifts to those produced by the magnetic sunspot model. The combination of longer path length and cooler plasma results in positive $\delta\tau$ for waves travelling away from the spot centre, where in the magnetic sunspot model we observed negative $\delta\tau$. A closer look at comparison plots in Figure $\ref{fig:1p5kG_thermal_times_plot}$ also reveals that the travel time increase due to fast-to-Alfv\'en conversion, typically seen around $\phi\approx80\degree-100\degree$, is absent in the thermal travel times. This indicates that the phase shifts produced by fast-to-Alfv\'en mode conversion are indeed distinguishable from thermal/geometrical effects and have a distinct and significant effect on helioseismic travel time measurements in sunspots. 

The measured thermal $\delta\tau$ at 3 and 5 mHz appear to reach their peak at $\theta \approx 32\degree$ in Figure $\ref{fig:1p5kG_thermal_times}$. However, we must remember that $\theta$ in the thermal calculations is purely representative of the source position, not actual field inclination from vertical, as magnetic effects are suppressed for these calculations. Hence, this apparent dependance on $\theta$ is a purely geometrical effect and is distinctly different from the ``ramp effect'' and fast-to-slow mode conversion-induced phase shifts we can observe in the magnetic $\delta\tau$ at $3$ and $5$ mHz in Figure $\ref{fig:1p5kG_times}$. 

\section{Discussion and Conclusions}

As solar imaging hardware becomes increasingly sophisticated, the need for innovative diagnostic tools and precise modelling of wave propagation and transformation properties in strong magnetic field regions of sunspots is made ever more apparent. Building on previous numerical studies which employed simple plane-parallel atmospheres, we conducted a non-exhaustive parametric study of waves in model sunspot atmospheres in an attempt to further our understanding of the implications of MHD mode conversion on helioseismic measurements. 

By using time-distance heliosiesmology and a travel-time measurement scheme sensitive to magnetic field orientation, we find that: i) The general behaviour of the travel-time shifts for the various sunspot models analysed is strikingly similar to that derived in \cite{cm2013} and \cite{mc2014}, being strongly linked to mode conversion in the atmosphere; ii) the magnitude of the directional travel times is dependant on the sunspot field strength, wave frequency and travel distance; iii) the depth of the Wilson depression can produce a measurable change in travel times, with slightly faster travel times produced by waves travelling in the direction of the sunspot axis as the Wilson depression depth is increased from $300$ to $500$ km below the surface; and finally iv) wave path changes produced by the underlying thermal structure of the sunspot appear to be the most significant contributor to the travel-time shifts for waves travelling towards and inside the umbra. Away from the umbra however, it is the magnetic effects that dominate.   

Overall, these results paint a fairly consistent picture: that the seismic waves' journey through the atmosphere can directly affect the wave travel times that are the basis of our inferences about the subsurface structure of sunspots, and in particular these effects are directional, depending on the orientation of the sunspot magnetic field. The close correspondence between these results and those derived previously using translationally invariant atmospheres, combined with the fact that directional filtering is directly extensible to real helioseismic data, argues strongly for the viability of directional time-distance probing of real solar magnetic regions. This will be the focus of future studies. 

In conclusion, directional helioseismology significantly enhances our computational helioseismology toolkit, where recently a number of other important advances have been made in both forward \citep{schunkeretal2013} and inverse \citep{hanasogeetal2012} modelling, ultimately leading to more precise helioseismic inferences of the subsurface structure and dynamics of sunspots. 
 
%%%%%%%%%%
%\section*{}
\vspace{2\baselineskip}\noindent
This work was supported by an award under the Merit Allocation Scheme on the NCI National Facility at the ANU, as well as by the Multi-modal Australian ScienceS Imaging and Visualisation Environment (MASSIVE). A portion of the computations was also performed on the gSTAR national facility at Swinburne University of Technology. gSTAR is funded by Swinburne and the Australian Government's Education Investment Fund. Dr Shelyag is the recipient of an Australian Research Council's Future Fellowship (project number FT120100057).

%%%%%%%%%%%%%%%%%%%%%%%%%%%%%%%
%  REFERENCES
%
%\bibliographystyle{nature}        
\bibliographystyle{mn2e}    
%\bibstyle{mn2e}    

%\bibliography{Biblio}

\end{document}